\let\texyear\year
\documentclass[hidelinks]{ieeeaccess}
\let\ieeeaccessyear\year
\let\year\texyear
\usepackage{orcidlink}
\usepackage{tikz}  
\definecolor{accessblue}{RGB}{0,105,154}
\let\year\ieeeaccessyear
\usepackage{cite}
\usepackage{amsmath,amssymb,amsfonts}
\usepackage{algorithmic}
\usepackage{graphicx}
\usepackage{textcomp}
\usepackage{multirow}
\usepackage{algorithm}
\usepackage{subfigure}
\usepackage{adjustbox}

\makeatletter
\newcommand\fs@norules{\def\@fs@cfont{\bfseries}\let\@fs@capt\floatc@ruled
  \def\@fs@pre{}%
  \def\@fs@post{}%
  \def\@fs@mid{\kern3pt}%
  \let\@fs@iftopcapt\iftrue}
\makeatother
\floatstyle{norules}
\restylefloat{algorithm}

\usepackage{tabularx,booktabs}  
\newcolumntype{C}{>{\centering\arraybackslash}X} 
\usepackage{lipsum} 

\def\BibTeX{{\rm B\kern-.05em{\sc i\kern-.025em b}\kern-.08em
    T\kern-.1667em\lower.7ex\hbox{E}\kern-.125emX}}

\begin{document}
\history{Date of publication xxxx 00, 0000, date of current version xxxx 00, 0000.}
\doi{10.1109/ACCESS.2017.DOI}

\title{Combining social relations and interaction data in Recommender System with Graph Convolution Collaborative Filtering}

\author{\uppercase{Tin T. Tran} \orcidlink{0000-0003-4252-6898}\authorrefmark{1},
\uppercase{Vaclav Snasel} \orcidlink{0000-0002-9600-8319}\authorrefmark{1},
\IEEEmembership{Senior Member, IEEE}, and \uppercase{Loc Tan Nguyen}\orcidlink{0009-0002-4926-398X}\authorrefmark{2}}
\address[1]{VSB-Technical University of Ostrava, 708 00 Ostrava, Czech Republic (e-mail: \{trung.tin.tran.st, vaclav.snasel\}@vsb.cz)}
\address[2]{Faculty of Information Technology, Ton Duc Thang University, Ho Chi Minh city, Vietnam (e-mail: 51900375@student.tdtu.edu.vn)}

\markboth
{Tin T. Tran \headeretal: Combining social relations and interaction data in Recommender System with GCCF}
{Tin T. Tran \headeretal: Combining social relations and interaction data in Recommender System with GCCF}

\corresp{Corresponding author: Tin T. Tran (e-mail: trung.tin.tran.st@vsb.cz).}

\begin{abstract}
A recommender system is an important subject in the field of data mining, where the item rating information from users is exploited and processed to make suitable recommendations with all other users. The recommender system creates convenience for e-commerce users and stimulates the consumption of items that are suitable for users. In addition to e-commerce, a recommender system is also used to provide recommendations on books to read, movies to watch, courses to take or websites to visit. Similarity between users is an important impact for recommendation, which could be calculated from the data of past user ratings of the item by methods of collaborative filtering, matrix factorization or singular vector decomposition. In the development of graph data mining techniques, the relationships between users and items can be represented by matrices from which collaborative filtering could be done with the larger database, more accurate and faster in calculation. All these data can be represented graphically and mined by today's highly developed graph neural network models. On the other hand, users' social friendship data also influence consumption habits because recommendations from friends will be considered more carefully than information sources. However, combining a user's friend influence and the similarity between users whose similar shopping habits is challenging. Because the information is noisy and it affects each particular data set in different ways. In this study, we present the input data processing method to remove outliers which are single reviews or users with little interaction with the items; the next proposed model will combine the social relationship data and the similarity in the rating history of users to improve the accuracy and recall of the recommender system. We perform a comparative assessment of the influence of each data set and calculation method on the final recommendation. We also propose and implement a model and compared it with base line models which include NGCF, LightGCN, WiGCN, SocialLGN and SEPT.
\end{abstract}


\titlepgskip=-15pt

\maketitle

\section{Introduction}

E-commerce develops strongly and gives users a good experience not only with complete information about products but also conveniences for customers such as providing items by catalog or comparing among products. Items selected according to certain criteria will be recommended to customers with relevant characteristics, such as age, gender, interests, or place of residence. Storing customer and product information is a challenging task because the number of customers and products is very large, increasing rapidly and diversifying sources of information collected. With a huge size database of users and items, the speed of access will limit the effectiveness of e-commerce. The collaborative filtering (CF) algorithm calculates the similarity between users \cite{5360986, 6779375}, without retrieving their personal information. This algorithm evaluates the similarity between customers based on the large number of items they have interacted with such as purchasing, rating, viewing, commenting. And then, the system will recommend to a target customer items in the interactive list of other customers who have a highly similarity with them. In these models, the correlation between the user and the item is recorded by an interaction matrix, which can be implicit or explicit. If the data are implicit, the matrix element will store whether the user is interested in or shopping for the item. If the data are explicit, the matrix element value represents the user's rating of the item. The similarity between two users is the distance between two vectors representing them, each of which contains information about that user's interactions with all items.

The explosion in the number of users and items leads to a huge matrix size, which will take up a large amount of memory for storage as well as computing resources for operation on matrices. The matrix factorization (MF) technique can decompose the interaction matrix into two or more matrices of smaller size without losing the original useful information. Singular value decomposition (SVD) algorithm generalizes the eigen decomposition of a square matrix which has an ortho normal eigen basis to any size matrix, and finally, the feature matrix is found with smaller size and minimum loss. However, most of the elements in these matrices have the value of zero, because almost users interested only a certain number of items and has no interaction with the rest of the item list. It made matrix sparse, they need to be compressed to both reduce the amount of memory needed as well as facilitate the computation. By matrix transformations, information is collected into embedding matrices and can be efficiently processed using TensorFlow and Keras libraries \cite{9321429}. As the final step of the computation, the embedding matrices can be converted back to the original interaction matrix size for the front-end layers to make recommendations to the user.

Besides exploiting user and item interaction information, users' social relationships are also essential because they show the user's influence in real life. A shopping recommendation from a friend will have a greater impact on a user's decision than other advertising sources. Today's social networking platforms such as Foursquare, Facebook, Gowalla also provide a database of users and their friendship relationships. Mining these facts will enrich the information and support not only the recommendation process but also specific user groups \cite{1458205, SoReg}.

Both the interactions between users and items and the social relationships between users can be graphically represented and mined using graph neural network (GNN) \cite{10068865}. In a graph, neighboring nodes in some hops show that they are related. Implicit information could be discovered by travel though the graph and look for nearly users who have the most common items reference. These high-order had been found in Neural graph collaborative filtering model (NGCF) \cite{NGCF}. Each matrix will also be transformed into embedded matrices that are significantly reduced in dimensions compared to the original matrix but still contain all the original information. The embedding matrices are decomposed to capture the influence signals and then combine them into a more informative embedding matrix \cite{Candillier2007ComparingSC}. The learning process can be repeated many times to reach the target accuracy, in the iterations propagated embedding matrices are used to capture high-order connectivity in the item and user interaction graph.

In this publication, we propose a GCN model that could analyst and synthesize from three sources of information: the product and user interaction matrix, the influence weighted matrix between users and the matrix of social friendships derived from social network platforms. We also implement the model and conduct experiments on well-known data sets and compare the results with the base line models. This article will be shown the next sections. Chapter 2 will summarize the research work that are the foundation of the topic and the latest techniques. Chapter 3 propose model in which we will show how to preprocess data sets, the organization of neural layers, and the signal aggregator and decomposer. Chapter 4 presents the overall results and compares them with previous state-of--the-art models. Chapter 5 is our conclusions and some ideas for future research.

\section{Literature Review}
\subsection{Collaborative Filtering}
Recommender system in the first stage exploited features of all items as well as the preferences of users to find suitable items and give reasonable recommendations to users \cite{Jalili18}. For example, a book can be categorized into comics, detective, historical, archaeology, fine arts, and others. With users, information such as age, gender, address of residence or education could also used to enrich the input data \cite{Pazzani99aframework}. However, with a very large number of items and frequent additions, it is difficult to find out and give items attributes, the systems need additive filtering methods. Collaborative filtering approach algorithms were used to find similarity between users or items without attribute information from them \cite{adomavicius2005incorporating}.

\Figure[t!](topskip=0pt, botskip=0pt, midskip=0pt)[width = 0.95\linewidth]{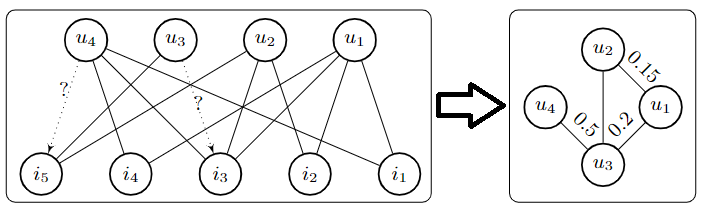}
{Collaborative Filtering algorithms calculates the correlation between users.\label{fig:correlation_user}}

Collaborative filtering can be implemented by memory-based models \cite{6313742} or model-based models \cite{10.1145/138859.138867}. With memory-based model, the history of every user's rating on items will be recorded in a rating matrix. There are many ways to define a rating scale, which can be an integer value from 1 to 5, or an implicit rating. In a space made up of user set, item set, and rating values, each user is represented by a feature vector $e_u$, also as each item is represented by feature vector $e_i$. The algorithms will find out the distance between each pair of users (or pair of items for item-based algorithms) to find the neighbors and form the recommended results. There are many methods to calculate distance such as Cosine function, Jaccard similarity or Mean Squared Differences \cite{Candillier2007ComparingSC}. 

In the model-based collaborative filtering systems, the algorithms will search for patterns in the learning data to develop a model for future prediction \cite{doi:10.1137/1.9781611972726.4}. The matrix that contains rating of users on items can be reduced by using MF techniques. In order to ensure accuracy of distance between users and items the rating matrix can be divided into user feature matrix and item feature matrix with smaller sizes \cite{5197422}. The higher accuracy of user similarity calculating, the better recommendation prediction results.

The Graph convolutional matrix completion (GCMC) built a graph-based auto-encoder framework for matrix completion \cite{berg2017graph}. This model moved from matrix to graph by defining the main problem as prediction task on a bipartite graph. Graph construction also benefits from having additional sources of information such as friendships in social networks, which are also represented by graphs. The encoders traverse the graph and record signals from vertices, representing users and items, producing latent features of these nodes. The of collaborative filtering problems between rows (or columns) has now become about measuring the predicted edge weights between two objects, which present users or items.

\subsection{Graph Convolution Networks}
A graph is a type of data structure that uses a set of vertices $V$ to represent an object and a set of edges $E$ to represent a relationship to depict the link between entities, people, data, and qualities. An edge on a graph may be directed or undirected, specifying the relationship's weight or just the relationship itself. In order to extract input data and identify properties of them, neural networks have had remarkable success setting up hidden layers \cite{9216015,gao2020deep}. In the publication \cite{wu2021graph}, neural networks were used with graph input data, and when vertex features were propagated and aggregated into one another during the learning process\cite{Wu2021}, favorable results were obtained. Graph neuron network (GNN) methods, which inherit from neural network algorithms, employ several propagation layers as well as various aggregation and update techniques. By means of pooling or attention techniques, the neighbor vertex's characteristics as well as those of the target vertex are better modified \cite{DBLP:journals/corr/abs-1903-07293}.

Graph Convolutional Network (GCN) is an iterative method of collecting information \cite{10.1145/3219819.3219890}. For example, when there is a similar item of interest to many of the target user's friends, it should be recommended with a higher degree than other items. GraphSAGE \cite{DBLP:journals/corr/HamiltonYL17} embedded inductive for each vertex of the graph and learned the topological structure of the graph as well as the effect of vertices on neighboring vertices. This method not only focus on feature-rich graphs but also make use of structural features of all vertices.

NGCF \cite{NGCF} generates hop-by-hop propagation classes in the input graph under the assumption that the effect of the users varies depending on the distance between the graph's vertices. Attention signals from a user's neighbors are received by the $k^{th}$ layer of propagation, also known as $k-order$ propagation, which also receives messages from items to that user. The input parameter $k$ can be viewed as the number of propagation layers, and the value $k=3$ is concluded to be the best one. The difference between the test set's actual and expected assessment value serves as the foundation for the loss function. Depending on the learning rate, the size of the embedding matrices, and the quantity of the learning data, the learning process to minimize the loss function occurs after a few iterations. LightGCN \cite{he2020lightgcn} deleted the feature weight transformation matrices during propagation and a non-linear activation function to remove adverse impacts on objects in the NGCF propagation process in a later version of the algorithm.

\subsection{Social Recommender system}
The social counseling system has been developed since the emergence of online social platforms. In addition to user behavior, restrictions such as liking, sharing and commenting; Additional data from social networks is used to provide personalized headlines to users considering the influence of their friends \cite{SoReg, 6714549 ,7404241}. Loads of information on social networks have influenced and made users of the same interests as their friends \cite{Bond2012, Qiu2018DeepInfM}.
ContextMF \cite{Wang2018ContextMFA} recorded the social relational information in the recommendation data in a frame design that was collected from matrix coefficients and with regularization terms. This model has improved the accuracy of the recommendation results when the model only performs MF with the user and item relation matrix. The TrustSVD model \cite{7404241} developed from SVD++ \cite{10.1145/1401890.1401944} incorporates the influence of friends on social relations as an implicit feedback add-on for an observing user .
GCN usage recommendation models are also extended to include more user social relation information, such as GraphRec \cite{fan2019graph} which viewed a user in both importance matrix aspects. social system and in the matrix of interactions with items. The data influence between users was aggregated and conditioned by the Social Aggregation module during machine learning.

SocialLGN \cite{LIAO2022595} designed deep models to capture the most distinctive characteristics of social and embedded networks and users of mixed embeddings, instead learning directly from their interaction data only users and items. Moreover, their model still uses GCN to study the reference of users whose information influence is affected through the process of social variables pervasive in social networks. User influence propagation also propagates through multiple logging layers, techniques used in NGCF \cite{NGCF} and Light GCN \cite{he2020lightgcn}.

Technically, the process of enhancement between social relational data and user influence in shopping is a challenge because they change according to the time and nature of each kind of items. In the Socially-Aware Self-Supervised Tri-training (SEPT) model \cite{DBLP:journals/corr/abs-2106-03569}, three histogram encoders were created to get the times signal from the social relationship between users, the user relationship when interested in similar items and information shared information - augment the data between social information and shopping interaction information. The self-supervised learning process in the above model has both improved the effectiveness of recommendations.

\section{Our proposed model}
In this section, we present a proposed model to exploit the rating matrix and social relationship between users to make recommendations for any user in the future. The model will be presented with data processing module, GCN module and prediction layer with loss calculator. An overview on our model show as Figure \ref{fig:overviewmodel}.


\Figure[t!](topskip=0pt, botskip=0pt, midskip=0pt)[width=1\textwidth]{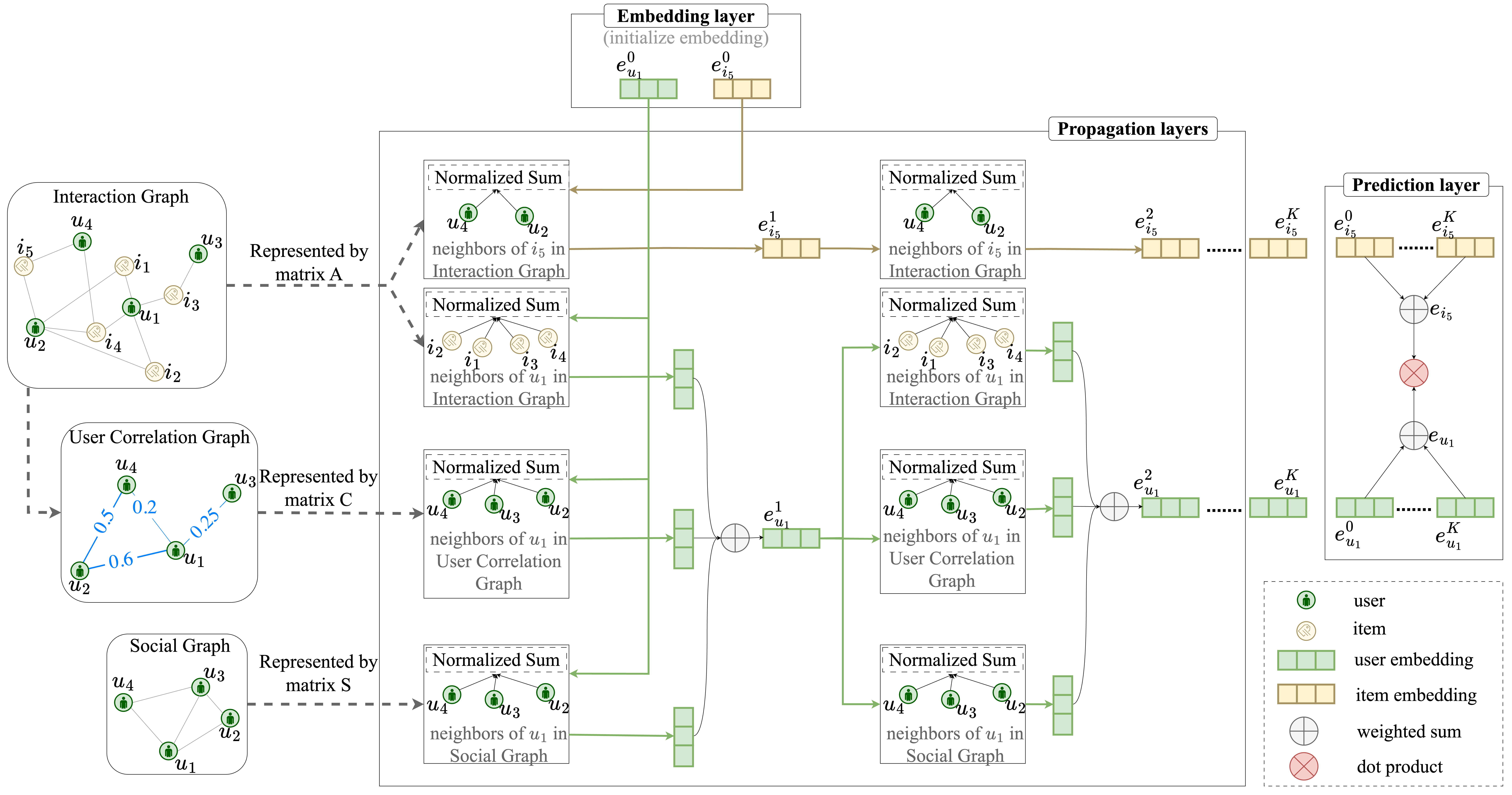}
{Overview on our proposed model.\label{fig:overviewmodel}}

\subsection{Data pre-processing}
The data sets are always recorded continuously and redundantly and included noise data that are users and items with very little interaction. Eliminating these signals will limit recommendations to items that are no longer available. Furthermore, the reduced data sets, smaller size but without losing valuable information, will decrease the memory size required, as well as cut down the computation time of the models. In our experiments, we apply a 10-core setting, which means that users with less than 10 interactions will be dropped at this processing step.

With the selected set of items, we also keep only the number of users with the highest interactions that are measured by the Jaccard distance between the set of items that a user interacts with and the set of all filtered items in the database. The ratio of number of users to number of items can be selected equal to the ratio of the original data set. The ratio of all survey data sets in the publication \cite{NGCF, he2020lightgcn}. We summarize this ratio in Table \ref{table:ratio}. Users with a low Jaccard measure were removed as outliers, and the social friendship matrix is being rebuilt accordingly.

\begin{table}
\centering
\caption{Ratio between number of users to number of items in surveyed datasets.}\label{table:ratio}
\begin{tabular}{|l|l|r|r|r|}
\hline
\textbf{Model} & \textbf{Dataset} &  \textbf{\#Users} & \textbf{\#Items} & \ \textbf{Ratio} \\
\hline
\multirow{3}{*}{NGCF} &
Gowalla & 29,858 & 40,981 & 1.37 \\
& Amazon-book & 52,643 & 91,599 & 1.74 \\
& Yelp2018 & 31,668 & 38,048 & 1.20 \\
\hline
\multirow{2}{*}{SocialLGN} &
Yelp & 17,237 & 38,342 & 2.22 \\
& Flickr & 8,358 & 82120 & 9.83 \\
\hline
\multirow{3}{*}{SEPT} &
Last.fm & 1,892 & 17,632 & 9.32 \\
& Douban-Book & 13,024 & 22,347 & 1.72 \\
& Yelp & 19,539 & 21,266 & 1.01 \\
\hline
\multirow{4}{*}{Our study} &
Gowalla & 28,511 & 40,968 & 1.44 \\
& Librarything & 9,988 & 15,985 & 1.60 \\
& Ciao & 5,785 & 108,651 & 18.78 \\
& Epinions & 1,497 & 17,898 & 11.95 \\
\hline
\end{tabular}
\end{table}

\begin{algorithm}
\caption{Filter out items with less than 10 interactions.}\label{alg:datafilter}
 \begin{algorithmic}[1]
 \renewcommand{\algorithmicrequire}{\textbf{Input:}}
 \renewcommand{\algorithmicensure}{\textbf{Output:}}
 \REQUIRE $\mathcal U \times \mathcal I $ \COMMENT{Interaction between users and items in original dataset}
\ENSURE R $\subseteq \mathcal U \times \mathcal I$ 
\FORALL{item $i \in \mathcal I $}
\IF{item $i$ has at least 10 interaction} \COMMENT{10-core setting}
    \STATE $I \gets \textrm{item } i$
\ENDIF
\ENDFOR
\STATE $p \gets \textrm{cardinality of set }I$
\STATE $q \gets p \div selected\_ratio$
\FORALL {user $u \in \mathcal U $}
\STATE $set_u \gets \textrm{list of items interacted by user u}$
\STATE $sim_u = \textrm{Jaccard distance between set $I$ and }set_u$
\ENDFOR
\STATE $U \gets q \textrm{ users have highest }sim_u$

\RETURN $R = U \times I$
\end{algorithmic}
\end{algorithm}

After pre-processing the data set, we've got 3 matrices as input to the model, where matrix A presents interaction between users and items, matrix C contains the similarity degree between users and matrix S represents social friendship of users. The matrix S is an optional matrix, which could be skipped if a data set does not provide a real social relationship; or when the noise of social relation data is higher than the actual effect among users.

\subsubsection{The interaction matrix A}
The user-item interaction data set can be provided as an implicit or an explicit. Implicit data records interactions such as item view, item purchase, mouse click, or like item as a binary value of 0 or 1, while explicit data specifies a user rating an item with a rating value in a certain range, for example an integer between 0 and 5. In this article, we consider the interactions between the user and the item are implicit recorded and expressed by a bipartite graph $G=\{V, E\}$, where vertices set $V=\{U, I\}$ is union from set of users $U$ and items set $I$; the E contains edge $e_{i,j}=(u_i, i_j)$ if user $u$ had interaction on item $i$. Thus, the matrix $R  \subseteq U  \times  I$, represents the graph $G$, could be defined by (\ref{eq:Rmatrix}).
    \begin{equation}
         R_{i,j}=\begin{cases}
            1 & \text{if user $u_i$ refers to item $i_j$ }\\
            0 & \text{otherwise}
         \end{cases} \label{eq:Rmatrix}
    \end{equation}

Interactive filtering by GCN will capture users' characteristics into user embedding while features of items will be learned into item embedding. In order for these two embedding matrices to be updated at the same time to facilitate propagation, we design the input matrix to be a Laplacian matrix showed in (\ref{eq:Aform}). If the matrix $R$ represents user interactions on the items, the transpose matrix $R^T$ will represent the items that the users interacted with. $0$ is a zero matrix of suitable size.

\begin{equation}
       A =  \begin{bmatrix}0 & R \\ R^ \top   & 0 \end{bmatrix} \label{eq:Aform}
\end{equation}

Because of convenience in calculation the embeddings in next layer in our graph convolution operations \cite{NGCF}, the symmetrically normalized matrix $\widetilde{A}$ should be calculated once by (\ref{eq:Anormal}) and accessed every time the embeddings are convoluted.
\begin{equation}\label{eq:Anormal}
     \widetilde{A} =  D^{ -\frac{1}{2}}AD^{ -\frac{1}{2}}
\end{equation}
where D is the diagonal degree matrix with each entry $D_{i, i}$ has value of number of nonzero entries in the $i-th$ row vector of degree matrix $A$ and $D_{i,j } = 0 $     with $i <> j $.
Finally, $\widetilde{A}$ is used as one of input sources of our GCN model.

\Figure[t!](topskip=0pt, botskip=0pt, midskip=0pt)[width = 0.95\linewidth]{IMG/WhyAform.png}
{The form of matrix M helps embeddings updated concurrency.\label{fig:formmatrix}}

\subsubsection{The users correlation matrix C}
The similarity between each pair of customers in the database depends on how many common items they interact with. Because the greater the number of items of common interest, the higher the influence between them. So that, in the publication \cite{wigcn}, a weighted user influence matrix $W_I$ was inputted as an additional source of information. This matrix can be calculated by $W_I = R \times R^ \top$ and it shows how many common items that user $u_i$ and $u_j$ have interaction by value of $ W_{I_{i,j}} $. On the other hand, $W_I$ also could be calculated by (\ref{eq:WI}).

However, the matrix $W_I$ only shows the number of intersections between the two sets of items $I_i$ and $I_j$, but has not recorded the influence of couple of users $\{i,j\} $ to all interaction data. The higher the total number of items both of customers has interacted with, the more similarity should be counted, since it is clear that they interact with the items more often than other customers. We propose the matrix $W_U$ has the same size as the matrix $W_I$ and has the following initialization value in (\ref{eq:WU}) and (\ref{eq:WI}) with where: $I_i$ and $I_j$ are the sets of items interacted by user $u_i$ and $u_j$, respectively.
\begin{equation}{\label{eq:WU}}
W_{U}=   |I_i| + |I_j| - |I_i  \cap  I_j| = |I_i  \cup  I_j| \end{equation}
\begin{equation}{\label{eq:WI}}
W_{I}=|I_i  \cap  I_j|
\end{equation}

Therefore, element-wise product $W_U  \odot {W_I}^{-1} $ will return a normalized matrix whose elements get a value in range from 0.0 to 1.0, based on the Jaccard similarity between each pair of users.

Furthermore, GCN-based models collect the collaborative signals from the high-order connectivity graph. The matrix $W_U  \odot {W_I}^{-1} $ inadvertently adds weight to the associations already extracted from high-order connectivity graph i.e. it makes the model more focused on the interaction points extracted from the high-order connectivity graph. This leads to a model that lacks extensible and is prone to over fitting. To solve this problem we use a function to convert value of elements of $W_U  \odot {W_I}^{-1} $ into the typical values which showed in Table \ref{table:classtifyfunction}.

\begin{table}
\centering
\caption{Represents the function $f$classified elements in the $W_U  \odot {W_I}^{-1} $ result matrix into denotation values.\label{table:classtifyfunction}}
\renewcommand\tabularxcolumn[1]{m{#1}}
\newcolumntype{P}[1]{>{\centering\arraybackslash}p{#1}}

\begin{tabularx}{1.0\linewidth}{|P{1.1 cm} |C|C|C|C|C|}

\hline

\textbf{Jaccard index} &  [0;0.1) & [0.1;0.4) & [0.4;0.6) & [0.6;0.9) & [0.9;1.0]\\
\hline
\textbf{Typical value} & 0.0 & 0.005 & 0.05 & 0.5 & 1.0\\
\hline
\textbf{Denote} & No interaction & Very low effective & Median effective & High effective & Strongly effective\\
\hline
\end{tabularx}
\end{table}

The conversion process of the resulting matrix $W_U  \odot {W_I}^{-1} $ also demonstrates the clustering of all users into multiple groups through their influence level, which is represented by the value of the matrix $C$.

Summarized all above ideas, the matrix C, which represents the degree of similarity between the behavior of the users, is implemented as the (\ref{eq:C}) with $f$ is the value allocation conversion function, shown in Table \ref{table:classtifyfunction}. $W_U$ is weighted user references matrix and $W_I$ is the matrix that represents the union of two lists of items two users.
\begin{equation}\label{eq:C}
    C = f(W_U  \odot {W_I}^{-1}) 
\end{equation}

\subsubsection{The social friendship matrix S}
Because the data of social relation is not always provided in a data set, it is an optional part of our recommendation system. In the next part of the experiment, we will present the comparative results with social relations and without it.
Some data sets and publications review the user $i$ and user $j$ relationships are represented as a directed edge in a graph if the user $i$ follow user $j$ ($j$ might not know about $i$). In this article, we treat the relationship of any couple of users as equal, or in other words, the edge of the relation between them in the graph is an undirected edge. The social relation matrix $S  \subseteq U  \times  U$ denotes the friendship among users in the real social life. If user $i$ and user $j$ are friends, $S_{ij} = 1$, otherwise $S_{ij} = 0$.

In recent studies, the matrix $S$ is used to enhance users-items relationships in input data. SEPT model has created a sharing view $A_s$\cite{DBLP:journals/corr/abs-2106-03569}, calculated as (\ref{eq:A_s}). $A_s$ represents the relationship weight between users not only by interesting in common items but also depending on whether they have a true relationship or not.
\begin{equation}\label{eq:A_s}
A_s = (RR^\top) \odot S
\end{equation}

We keep our matrix $S$ separate from interaction data $A$ and users correlation values $C$ in the input to evaluate the influence of friendship relations on overall recommendation accuracy. We leave research on integrating social signals into other matrices for future research.

\subsection{Graph Convolution Network layers}
\subsubsection{Transformation weight matrices and activation function}
The NGCF model represents the most advanced models that use GCN for mining and recommendation. NGCF is designed with feature transformation matrices as well as a nonlinear activation function like other GCN standards \cite{najork2008computing}. Then the non-linear activation will incorporate embedding in the output and smoothing the values so as not to introduce a biased value in overall result. The LeakyReLU function is commonly used in GCN models \cite{NGCF, wigcn}.

However, developing the NGCF model on data with little description, where both user entries and each other are stored with only the valid ID, does not bring any benefit to the feature learning process. On the contrary, computer resources are spent a lot on storage and operations on digital matrices. The use of weighted transformation matrices and activation function has been discussed in \cite{he2020lightgcn}. In this publication, the LightGCN model eliminates all the weighted transformation $W_i$ along with the activation function and achieves better overall results than the NGCF model.
We again test this method on the SocialLGN model \cite{LIAO2022595} with the variants eliminating either the transformation matrices, the function activations, or both.
\begin{itemize}
    \item \textbf{SocialLGN-f} removes all feature transformation matrices $W_1, W_2$ and $W_3$.
    \item \textbf{SocialLGN-n} eliminates the non-linear activation function $\sigma(.)$.
    \item \textbf{SocialLGN-fn} eliminates both the feature transformation matrix $W_i$ and the nonlinear activation function $\sigma(.)$.
\end{itemize}

\begin{figure}[H]
    \centering
    \includegraphics[width = 0.9\linewidth]{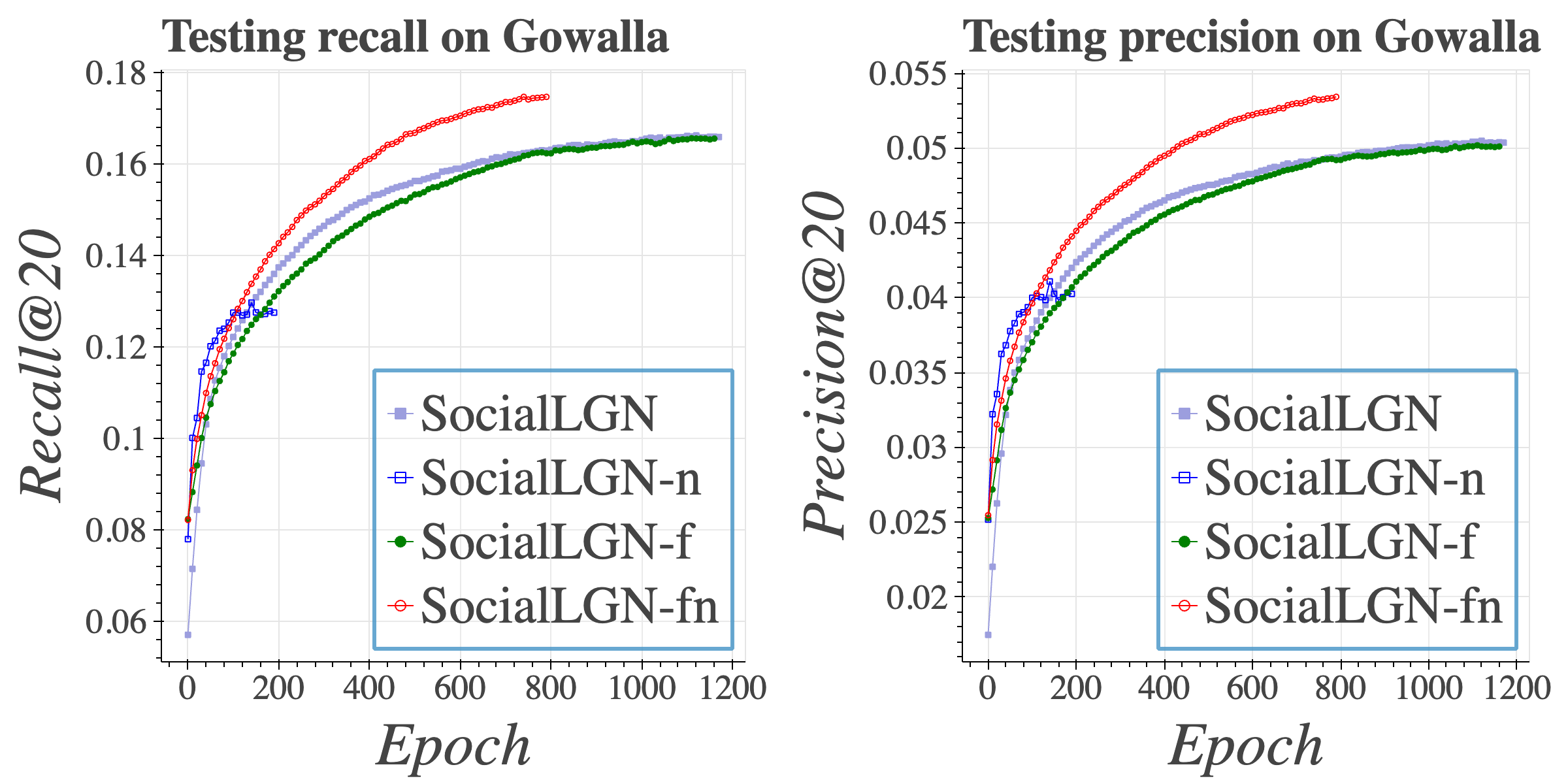}
    \caption{\centering Compare variants of SocialLGN on data set Gowalla.
    \label{fig:compareGowalla}}
\end{figure}

\begin{figure}[H]
    \centering
    \includegraphics[width = 0.9\linewidth]{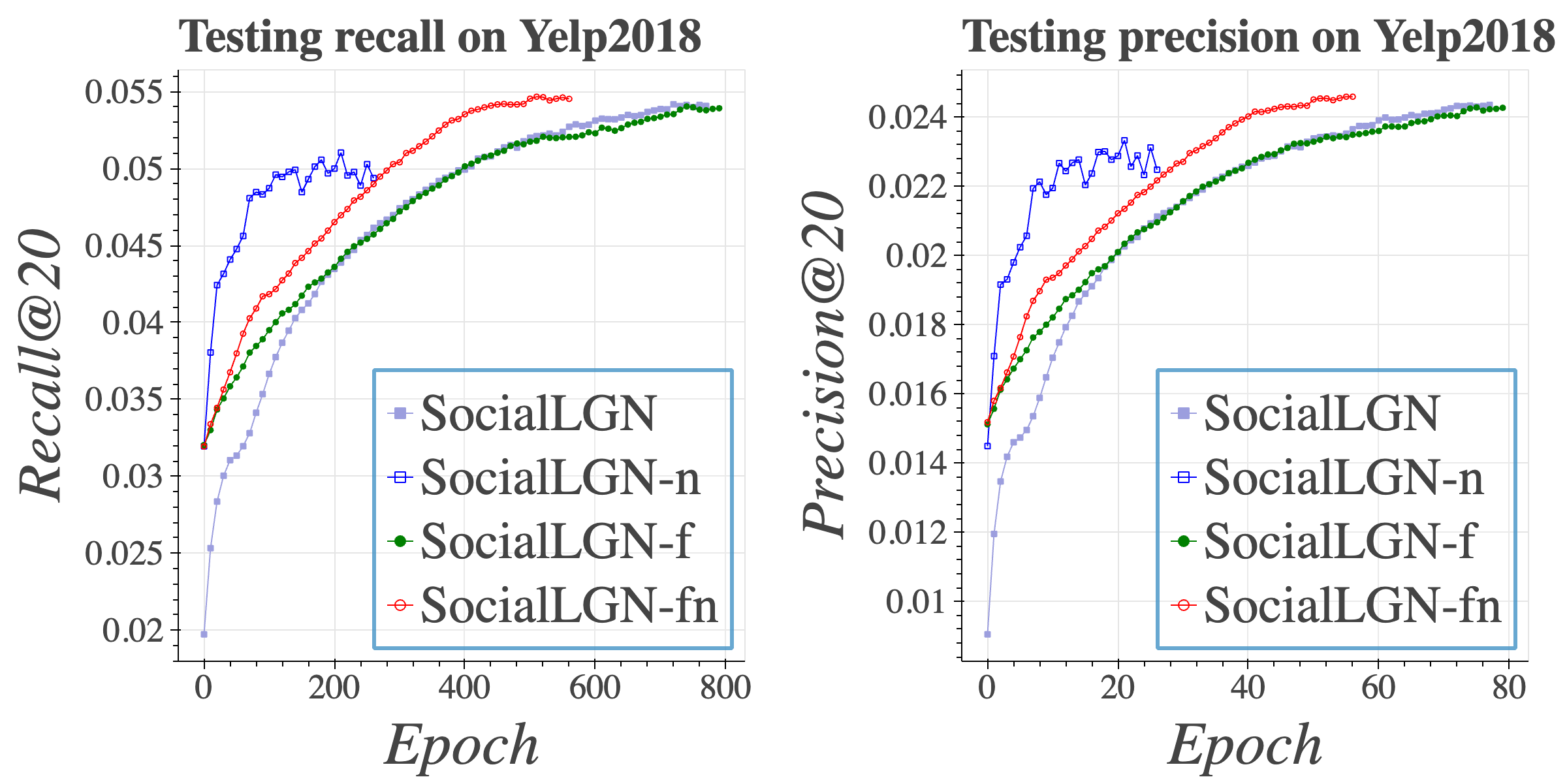}
    \caption{\centering Compare variants of SocialLGN on data set Yelp2018.\label{fig:compareYelp}}
\end{figure}

In all experiments, we apply the same hyperparameters with publications NGCF and LightGCN, including learning rate, regularization, dropout rate, Gowalla and Yelp2018 data. The results show in Figure \ref{fig:compareGowalla} with Gowalla data set and Figure \ref{fig:compareYelp} with Yelp data set.

When removing either feature transformation matrices $W_1,W_2,W_3$ or non-linear activation functions $\sigma (.)$, the model performance is unstable and worse than the original SocialLGN, but removing both feature transformation matrices and activation function the model improves significantly. Through this experiment we proved our inference about the negative impact of graph fusion operation to embedding users. To replace the graph fusion operator, at SocialLGN-fn we simply sum the embedding users social and the embedding users interaction together to generate embedding users for each propagation.

From the above presentation, we determine that eliminating both the feature transformation matrices and the nonlinear activation function is appropriate for the data sets considered in this article.

\subsubsection{Embedding layers}
Follow the model in publication \cite{he2017neural} used two vectors for user's latent and item's latent, to define as a result are users embeddings and items embeddings, we also define users embeddings and items embeddings as $e_u  \in \mathbb{R}^d$ and $e_i \in \mathbb{R}^d$ with $d$ is the embeddings vector size. In the first round of propagation, embedding layer $e^{(0)}=e_u^{(0)} \parallel e_i^{(0)}$ will be initialized with normal weight initialization method. Based on the architecture of LGC \cite{he2020lightgcn}, the initialized embeddings $e_u^{(0)}$ and $e_i^{(0)}$ were the only trainable parameters of our model. In matrix form we denote $E^{(0)}\in{\mathbb{R}^{(n+m) \times d}}$ is the set of all embeddings during propagation, i.e. $E^{(0)}$ contains the set of $n$ user embeddings  and $m$ item embeddings.

\begin{equation}
    E^{(0)}=E_U^{(0)} \parallel E_I^{(0)}=[e_{u_1}^{(0)},\dotso ,e_{u_n}^{(0)},e_{i_1}^{(0)},\dotso,e_{i_m}^{(0)}]
\end{equation}


\subsubsection{Propagation process}
In order to capture signals from three input matrices, every $k$ propagation layers we propagate the $k^{th}$ user and item embeddings $e^k_u$ and $e^k_i$ on the LGC architecture \cite{he2020lightgcn} and obtain four embeddings containing three signals are $e_{ua}^{(k+1)}, e_{c}^{(k+1)}, e_{s}^{(k+1)}, e_{i}^{k+1}$. Where $e_{ua}^{(k+1)}$ and  $e_{i}^{(k+1)}$ hold user-item interaction signals in the matrix $A$, $e_{s}^{(k+1)}$ contain social signals between users in matrix $S$, and $e_{c}^{(k+1)}$ capture correlation signals between users in matrix $C$.

\begin{equation} \label{eq:cal_embedding}
    \begin{split}
        e_{ua}^{(k+1)} &=\sum_{i\in \mathcal{N}_u^A}^{}\frac{1}{ \sqrt{| \mathcal{N}_u^A ||\mathcal{N}_i^A| } }e_{i}^{(k)}\\
        e_{s}^{(k+1)} &=\sum_{s\in \mathcal{N}_u^S}^{}\frac{1}{ \sqrt{| \mathcal{N}_u^S ||\mathcal{N}_s^S| } }e_{u}^{(k)}\\
        e_{c}^{(k+1)} &=\sum_{c\in \mathcal{N}_u^C}^{}\frac{1}{ \sqrt{| \mathcal{N}_u^C ||\mathcal{N}_c^C| } }e_{u}^{(k)}\\
        e_{i}^{(k+1)} &=\sum_{u\in \mathcal{N}_i^A}^{}\frac{1}{ \sqrt{| \mathcal{N}_i^A ||\mathcal{N}_u^A| } }e_{u}^{(k)}
    \end{split}
\end{equation}

where $|\mathcal{N}_q^X|$ denote the number of neighboring users (or items) of item(or user) $q$ in the matrix $X$, $X = [A, S, C]$. The essence of equation (\ref{eq:cal_embedding}) is based on the symmetric normalization element was used in most GCN model \cite{NGCF, he2020lightgcn, wigcn} and was calculated by (\ref{eq:Anormal}).

Then the $(k + 1)^{th}$ user embeddings are aggregated according to the equation (\ref{eq:cal_user}), specifically, we combine three embedding by using the weighted sum of the embeddings.

\begin{equation} \label{eq:cal_user}
e_{u}^{(k+1)} =AGGREGATION_k \left (  e_{ua}^{(k+1)},\quad e_{c}^{(k+1)}, \quad e_{s}^{(k+1)} \right )
\end{equation}

After some loops of propagation process, we've got $(K+1)$ embeddings corresponding to the number of propagations through $K$ layers and including the initial embedding. The final embedding of users (and items) will be obtained by (\ref{eq:final_embedding}).

\begin{equation} \label{eq:final_embedding}
e_u = \frac{1}{K}\sum_{k=0}^{K}e_u^{(k)};\quad e_i = \frac{1}{K}\sum_{k=0}^{K}e_i^{(k)}
\end{equation}

Using equation (\ref{eq:final_embedding}), we takes the mean value of all embeddings at all layers, and that equation can be replaced with some other functions like maximum, median, weighted average. In order not to complicate the model and still ensure good performance in general, we use the mean function.

\subsubsection{Prediction and optimization}
Based on the final embedding $e_u$ and $e_i$ calculated by (\ref{eq:final_embedding}), we calculate the prediction score of item $i$ for user $u$ by equation (\ref{eq:cal_prediction}).
\begin{equation} \label{eq:cal_prediction}
    \widehat{y}_{ui}=e_{u}^\top  e_{i}
\end{equation}

We build the loss function with Bayesian personalize ranking (BPR) for model to learn the parameters $\Phi$ which which only include the user and item embedding. BPR is the best suitable method for implicit feedback data sets \cite{rendle2012bpr}, it assumes observed interactions $\Omega^+_{ui}$ have higher preferences than an unobserved interactions $\Omega^-_{uj}$. To optimize the prediction model we use mini-batch Adam \cite{kingma2017adam} and minimize the  BPR loss in (\ref{eq:loss_cal}).
\begin{equation} \label{eq:loss_cal}
    Loss_{BPR} =  \sum_{\Omega^+_{ui}}\sum_{\Omega^-_{uj}} -\ln \sigma ( \widehat{y}_{ui} - \widehat{y}_{uj} ) +  \lambda  \parallel  \Phi   \parallel ^2_2
\end{equation}

\subsubsection{Matrix Form}
To facilitate the implementation of the model, we expressed the propagation process in the form of matrix in (\ref{eq:Enext}).
\begin{equation}
\begin{split}
    E^{(k+1)} &= E_U^{(k+1)} \parallel E_I^{(k+1)}\\
    &= (E_{U_{A}}^{(k+1)} + E_S^{(k+1)} + E_C^{(k+1)}) \parallel E_I^{(k+1)}\\
    &= (\widetilde{R}E_I^{(k-1)} + \widetilde{S}E_U^{(k-1)} + \widetilde{C}E_U^{(k-1)}) \parallel \widetilde{R^\top}E_U^{(k-1)}
\end{split}\label{eq:Enext}
\end{equation}
where $\widetilde{C}$ and $\widetilde{S}$ is a symmetrically normalized matrix of $C$ and $S$ as in (\ref{eq:sym}).

\begin{equation}
\begin{split}
    \widetilde{S} &= D_{S}^{-\frac{1}{2}}SD_{S}^{-\frac{1}{2}}\\
    \widetilde{C} &= D_{C}^{-\frac{1}{2}}CD_{C}^{-\frac{1}{2}}
\end{split}\label{eq:sym}
\end{equation}

$\widetilde{R}$ and $\widetilde{R^\top}$ appeared as components in the input $\widetilde{A}$ in (\ref{eq:Anormal}).

\section{Results}
\subsection{Experimental Setting}
We use setting 10-core with all data sets to ensure that each user has at least ten interactions. For each data set, we selected 20\% of interactions which have the latest timestamp for testing set and the remaining (80\% interactions) are for process of training. Summary of all data sets showed in Table \ref{table:Static}.
We configure our model with embedding fixed size to 64 for all models and the embedding parameters are initialized with the normalized weight method. The optimization process was done with Adam algorithm \cite{kingma2017adam}. 
\subsubsection{Data sets}
\begin{itemize}
    \item \textbf{Gowalla}: First published in \cite{10.1145/2020408.2020579}, Gowalla is a web-based application that provides location-based social networking. Users can check-in and share public places. Gowalla also collects the friendship network of users and provides these data as an undirected graph.
    \item \textbf{Librarything}: is a data from a book review website called Library Thing \cite{10.1145/2806416.2806511}. It is an online service to help people organize their books easily and people can communicate with each other. 
    \item \textbf{Ciao}: is an online shopping site that records users’ ratings of items with timestamps. Customers can rate an item with a score from 1 to 5. Users can also add others to their friend lists and build a social network \cite{10.1145/2124295.2124309}.
    \item \textbf{Epinions}: is a well-known consumer review site where users can rate items and add social friends to their trust lists \cite{epinionds}. The Ciao and Epinions data sets have been widely selected as benchmark data sets for social recommendations. 
\end{itemize}

\begin{table}
\centering
\caption{Statistic of the experiment data sets.}\label{table:Static}
\renewcommand\tabularxcolumn[1]{m{#1}}
\begin{tabularx}{0.9\linewidth}{|C|r|r|r|r|}
\hline
\textbf{Data set} & \textbf{ \#Users} & \textbf{\#Items} & \textbf{\#Edges} & 
\textbf{\#Social}\\
\hline
Gowalla & 28,551 & 40,968 & 992,860 
& 265,704\\
Librarything & 9,988 & 15,985 & 372,967 
& 28,736\\
Ciao & 5,785 & 108,651 & 283,034 
& 101,245\\
Epinions & 1,497 & 17,898 & 25,191 
& 9,493\\
\hline
\end{tabularx}
\end{table}

\subsubsection{Evaluation criteria}
The evaluative measurement should be selected appropriately for each algorithm \cite{10.1145/963770.963772}. A commonly evaluating method is dividing a data set into a test set and a train set. The algorithm is applied on the train set to make predictions and evaluated these results on the test set. The difference between the learning result and the actual data value shows the accuracy of the algorithm of proposed model. This difference can be represented in mean absolute error (MAE) and root mean square error (RMSE). Besides accuracy, recall, scalability, learning time, memory consumption or interpretability are also an important criterion in evaluating the recommended system.

For implicit data, interactions between the user and the item are recorded binary, rather than rated as a specific value. Then, the algorithms will use the accuracy measure for the classification. The commonly used metrics are Precision and Recall \cite{Sarwar00, Sarwar00E}. Precision is the ratio between correct predictions on the test set, and Recall is the sensitivity of the algorithm, or the proportion of relational assertions that have been retrieved from the test set.  
\begin{equation} \label{eq:precision}
    \begin{split}
        Precision &= \frac{TP}{TP+FP} \\
        Recall &= \frac{TP}{TP+FN} \\
    \end{split}
\end{equation}
where $TP$ is true positive set of correctly predicting interaction exists between users and items. $FP$ is false positive set of missing prediction of interaction while $FN$ (false negative) presents the number of predicted interactions not being exist on the test set.

Furthermore, Discounted Cumulative Gain score (DCG) \cite{najork2008computing} assumes that judges have assigned labels to each result and accumulates across the result vector a gain function $G$ applied to the label of each result, scaled by a discount function $D$ of the rank of the result, and it's normalized by the dividing DCG of an ideal result vector $I$. We get Normalized Discounted Cumulative Gain (NDGC) as in (\ref{eq:ndgc}).
\begin{equation} \label{eq:ndgc}
        NDGC_u = \frac{DCG_u}{CDG_{max}}
\end{equation}
In this publication, we use Precision and Recall (\ref{eq:precision}) and NDCG (\ref{eq:ndgc}) that considered at 20 items (NDGC@20).
\subsubsection{Base line models}
To evaluate our proposed model, we compare it with the following state-of-the-art models as base lines.
\begin{itemize}
    \item \textbf{MF}\cite{5197422} is the matrix factorization model demonstrated in the Netflix pricing competition. The model outperforms classic nearest neighbor techniques for generating product recommendations and allows the inclusion of additional information such as implicit feedback, temporal effects, and confidence. Even though this model is outdated, we still put it in comparison to show the evolution of recommendation models.
    \item \textbf{Social MF}\cite{10.1145/1864708.1864736} incorporates the trust information and its propagation into an MF model. The signals from direct friends are embedded in the final matrix. 
    \item \textbf{Trust SVD}\cite{7404241} is a trust-based MF technique, is built on the SVD++ algorithm. This model handles the explicit and implicit influence of rated items, by incorporating reviews from trusted friends, to make prediction of items for an active user.
    \item \textbf{GCMC}\cite{berg2017graph} assumes that making a future recommendation is a prediction of the association between the user and the item in the graph. Interaction data is represented by one bipartite graph with labeled edges denotes the observed ratings. The model then uses a graph auto encoding framework based on message passing on interactive graphs.
    \item \textbf{NGCF}\cite{NGCF} is one of the most cutting-edge GCN models. It performs propagation operations on embeddings using a few iterations. High-order connectivity in the interactions graph is contained in the stacked embeddings on the output. The latent vectors contain the collaborative signal, which strengthens the model.
    \item \textbf{WiGCN}\cite{wigcn} based on NGCF model, the WiGCN added a weighted matrix as an additional input. That matrix contains the influence of users on each other. It leads to more data-gathering propagation and increased recommendation performance.
    \item \textbf{LightGCN}\cite{he2020lightgcn} in order to concentrate on the neighborhood aggregation element for CF, this model eliminates the weight matrices and the activation function. The users and items embeddings for the interaction graph are learned using linear propagation in this model. The weighted total of all learnt embeddings becomes the final embedding.
    \item \textbf{SocialLGN}\cite{LIAO2022595} concentrate on the CF's neighborhood aggregation component. The users and items embeddings for the interaction graph are learned using linear propagation in this model. The weighted total of all learnt embeddings becomes the final embedding.
    \item \textbf{SEPT}\cite{DBLP:journals/corr/abs-2106-03569} from the user social information, this model augments the user data views with the user social information. And then the framework builds three graph encoders upon the augmented views and iteratively improves each encoder with self-supervision signals from other two encoders.
\end{itemize}
\begin{table*}[t]
\caption{Overall Performance Comparisons}\label{tab:Result}
\renewcommand\tabularxcolumn[1]{m{#1}}
\begin{tabularx}{1.0\textwidth}{|c|CCC|CCC|CCC|CCC|}
\hline
 \textbf{Data set} & \multicolumn{3}{c|}{\textbf{Gowalla}} & \multicolumn{3}{c|}{\textbf{Librarything}} & \multicolumn{3}{c|}{\textbf{Ciao}} & \multicolumn{3}{c |}{\textbf{Epinion}}\\
   & recall & precision & ndcg* &  recall & precision & ndcg* & recall & precision & ndcg* & recall & precision & ndcg* \\ 
\hline
MF &  0.1091 &  0.0308 &  0.0920 &  0.0472 &  0.0114 &  0.0348 & 0.0318 & 0.0105 & 0.0222  & 0.0049 & 0.0006 & 0.0026\\
SocialMF & 0.1167 & 0.0327 & 0.0980 & 0.0493 & 0.0121 & 0.0367 & 0.0348 & 0.0103 & 0.0241 & 0.0074 & 0.0009 & 0.0036  \\
TrustSVD & 0.1376 & 0.0380 & 0.1120 & 0.0539 & 0.0137 & 0.0374 & 0.0366 & 0.0107 & 0.0248 & 0.0067 & 0.0008 & 0.0035 \\
GCMC &  0.1357 &  0.0385 &  0.1097 &  0.0571 &  0.0151 &  0.0366 & 0.0466 & 0.0139 & 0.0334 & 0.0075 & 0.0009 & 0.0036 \\
NGCF &  0.1622 &  0.0445 &  0.1299 &  0.0680 &  0.0166 &  0.0442 & 0.0506 & 0.0149 & 0.0353 & 0.0118 & 0.0015 & 0.0065 \\
WiGCN &  0.1632 &  0.0445 &  0.1321 &  0.0734 &  0.0173 &  0.0455 & 0.0522 & 0.0153 & 0.0369 & 0.0122 & 0.0015 & 0.0058 \\
LightGCN &  0.1808 &  0.0496 &  0.1488 &  0.0768 &  0.0185 &  0.0507 & 0.0571 & 0.0168 & 0.0431 & 0.0150 & 0.0022 & 0.0085 \\
SocialLGN &  0.1705 &  0.0467 &  0.1411 &  0.0741 &  0.0180 &  0.0509 & 0.0521 & 0.0158 & 0.0384 & 0.0102 & 0.0015 & 0.0060  \\
SEPT &  0.1872 &  0.0525 &  0.1519 &  0.0763 &  0.0184 &  0.0499 & 0.0586 & 0.0176 & 0.0439 & 0.0138 & 0.0018 & 0.0071 \\
\hline
Our w/ interact   &  0.1846 &  0.0506 &  0.1509 &  0.0770 &  0.0188 &  0.0509 & 0.0592 & 0.0173 & 0.0433 & 0.0139 & 0.0019 & 0.0073  \\
Our w/ social   &  0.1924 &  0.0530 &  0.1564 &  0.0774 &  0.0185 &  0.0498 & 0.0578 & 0.0170 & 0.0439 & 0.0100 & 0.0019 & 0.0055  \\
\textbf{Our model-all}   &  \textbf{0.1940} &  \textbf{0.0537} &  \textbf{0.1572} &  \textbf{0.0804} &  \textbf{0.0191} &  \textbf{0.0533} & \textbf{0.0594} & \textbf{0.0176} & \textbf{0.0444} & \textbf{0.0175} & \textbf{0.0023} & \textbf{0.0096} \\
\hline
\end{tabularx}
\noindent{\footnotesize{* ndcg@20}}
\end{table*}
\subsection{Overall result comparison}

Summary of all results from all models on the processed data are displayed in the Table \ref{tab:Result} with three values in each data set: precision, recall and ndcg measured in 20 items. In order from the top row down, the chemical process of recommender system models can be clearly identified, that is, the group of methods using GCN has better results than the group of classical methods such as MF or SVD. In the group of GCN methods, models that do not use the transformation matrices, such as LightGCN, give better results than the remaining models. When compared in the group of models that do not consider social relationships, that is MF, GCMC, NGCF, LightGCN, WiGCN, our model with interaction embedding (\textbf{Our w/ interaction}) still gives the highest results.

In each pair of models such as MF and Social MF, LightGCN and SocialLGN, LightGCN and SEPT, almost models that use social relationship give better results; This confirms that mining actual friend data are as important as mining the correlation between users through items of common interest. Our proposed model also efficiently accounts for social relationship data when embedding them in the propagation process.
\subsection{Detailed Model Analysis}
\subsubsection{Effect of social relation on overall result}
We compare each pair of models that are architecturally similar to each other in Table \ref{table:socialimprovement}. In each row of comparison, one model exploits the users and items relationship data while the other model adds users and users relationship data. In a pair of models, the signal propagation and reception formulas all have a high degree of similarity, using (or not using) the weighted transformation matrices and non-linear activation functions. The parameters setting in the experiment are the same.
\begin{table}[H]
\centering
\caption{Improvement with social friendship data on Gowalla data set.\label{table:socialimprovement}}
\renewcommand\tabularxcolumn[1]{m{#1}}
\newcolumntype{P}[1]{>{\centering\arraybackslash}p{#1}}
\begin{tabularx}{1.0\linewidth}{|P{1.7cm} |P{1.7cm}|C|C|C|}
\hline
\textbf{Users-Items models} & \textbf{Social models} & \textbf{recall} & \textbf{precision} & \textbf{ndcg@20}\\
\hline
MF & SocialMF & $\uparrow 7.0\% $ & $\uparrow 6.2\% $ & $\uparrow 6.5\% $\\
LightGCN & SocialLGN & $\downarrow 5.7\% $ & $\downarrow 5.8\% $ & $\downarrow 5.2\% $\\
LightGCN & SEPT & $\uparrow 3.5\% $ & $\uparrow 5.8\% $ & $\uparrow 2.1\% $\\
Our w/ interaction & Our model-all & $\uparrow 5.1\% $ & $\uparrow 6.1\% $ & $\uparrow 4.2\% $\\

\hline
\end{tabularx}
\end{table}

\begin{enumerate}
    \item From MF to SocialMF: in the traditional MF model, the characteristics of each user have been obtained from the decomposition of the users - items relationship matrix. Meanwhile, the SocialMF model also extracts characteristics of a user's direct friends to enrich information for that user. However, the SocialMF model did not deepen the indirect relationships between the user community, so it missed many valuable signals.
    \item From LightGCN to SocialLGN: this is the only case where the results are worse across all measures. The process of combining signals from mining similarities between users with the social relationships of users is not in harmony, resulting in friend information not only reducing the accuracy of the recommendation results but also scattering the CF results.
    \item From LightGCN to SEPT: although both SocialGCN and SEPT are models that extend from LightGCN with friendship information, it is clear that the SEPT model has optimized the embedding of the friendship matrix between users into the model and brings the benefits big improvement.
    \item We also tested the influence of friendship data on the overall results by removing the S matrix in the model input and comparing the results between the model without the S matrix (Our model with only interaction data) and full model (Our model-all). We calculated the influence weights between users and clustering gathered those weights into each relationship level as shown in the Table \ref{table:classtifyfunction}. By doing that way, our model avoided the damage of SocialLGN that still captures valuable friend signals like in the SocialMF model and explores indirect relationships like other GCN models.
\end{enumerate}
Although friendship data has been shown to influence the final recommendation results, it should be noted that relationships in real world change over time and can introduce bias in recommendations. An example is that friends in the same type of hobby will find it difficult to give good advice about another type of hobby.

\subsubsection{Time performance analysis}
The time cost of our model lies in the interaction and social relation graph aggregation. Therefore, the total time complexity is acceptable in practice. We try to make the environment run the experiments as similar as possible each time, the average time results of each epoch compared to each data set on the LGC-based models are almost similar, the figures statistics are given in Table \ref{table:time_experiments}.

\begin{table}[H]
\newcolumntype{P}[1]{>{\centering\arraybackslash}p{#1}}
\caption{The number of epochs the models learned and the average time (in seconds) per epoch.}
\label{table:time_experiments}
\begin{adjustbox}{width=\columnwidth,center}

\begin{tabular}{|c|c|c|c|P{1.1cm}|P{1.1cm}|P{1.1cm}|}

\hline
                  & \textbf{WiGCN} & \textbf{LightGCN} & \textbf{SEPT} & \textbf{Our w/ interact} & \textbf{Our w/ social} & \textbf{Our model-all} \\ \hline
\textbf{Gowalla} &
  \begin{tabular}[c]{@{}c@{}}350\\ (58.2s)\end{tabular} &
  \begin{tabular}[c]{@{}c@{}}1140\\ (20.4s)\end{tabular} &
  \begin{tabular}[c]{@{}c@{}}1090\\ (22.8s)\end{tabular} &
  \begin{tabular}[c]{@{}c@{}}490\\ (22.5s)\end{tabular} &
  \begin{tabular}[c]{@{}c@{}}1040\\ (22.8s)\end{tabular} &
  \begin{tabular}[c]{@{}c@{}}440\\ (23.7s)\end{tabular} \\ \hline
\textbf{LibraryThing} &
  \begin{tabular}[c]{@{}c@{}}380\\ (32.3s)\end{tabular} &
  \begin{tabular}[c]{@{}c@{}}980\\ (12.3s)\end{tabular} &
  \begin{tabular}[c]{@{}c@{}}970\\ (13.1s)\end{tabular} &
  \begin{tabular}[c]{@{}c@{}}350\\ (13.3s)\end{tabular} &
  \begin{tabular}[c]{@{}c@{}}990\\ (13.3s)\end{tabular} &  
  \begin{tabular}[c]{@{}c@{}}570\\ (13.5s)\end{tabular} \\ \hline
\textbf{Ciao} &
  \begin{tabular}[c]{@{}c@{}}70\\ (25.8s)\end{tabular} &
  \begin{tabular}[c]{@{}c@{}}240\\ (9.1s)\end{tabular} &
  \begin{tabular}[c]{@{}c@{}}220\\ (10.2s)\end{tabular} &
  \begin{tabular}[c]{@{}c@{}}180\\ (9.7s)\end{tabular} &
  \begin{tabular}[c]{@{}c@{}}260\\ (10.1s)\end{tabular} &
  \begin{tabular}[c]{@{}c@{}}160\\ (10.2s)\end{tabular} \\ \hline
\textbf{Epinions} & 
  \begin{tabular}[c]{@{}c@{}}50\\ (12.5s)\end{tabular} &
 \begin{tabular}[c]{@{}c@{}}170\\ (4.2s)\end{tabular} &
  \begin{tabular}[c]{@{}c@{}}160\\ (4.8s)\end{tabular} &
  \begin{tabular}[c]{@{}c@{}}90\\ (5.3s)\end{tabular} &
    \begin{tabular}[c]{@{}c@{}}170\\ (4.3s)\end{tabular} &
  \begin{tabular}[c]{@{}c@{}}140\\ (4.9s)\end{tabular} \\ \hline        
\end{tabular}
\end{adjustbox}
\end{table}
An extremely outstanding feature of our proposed model is its convergence speed. Not only does it have higher recommendation predictions than other models, but its time efficiency is also significantly shorter than LGC architecture-based models such as LightGCN, SocialLGN, and SEPT. Specifically, in the Gowalla data set, our model with only interaction data takes 490 epochs and Our model only takes 440 epochs, while the two models SEPT and LightGCN take 1090 and 1140 epochs, respectively. For the remaining two data sets, Ciao and Epinions, there is also a similar trend, which demonstrates very good performance of the model training process. We conclude that the user correlation matrix helps the model focus on important data, thereby accelerating the learning process of our model.

\subsubsection{Positive impact of user interaction matrix}
\begin{figure}[H]
    \centering
    \includegraphics[width = 0.9\linewidth]{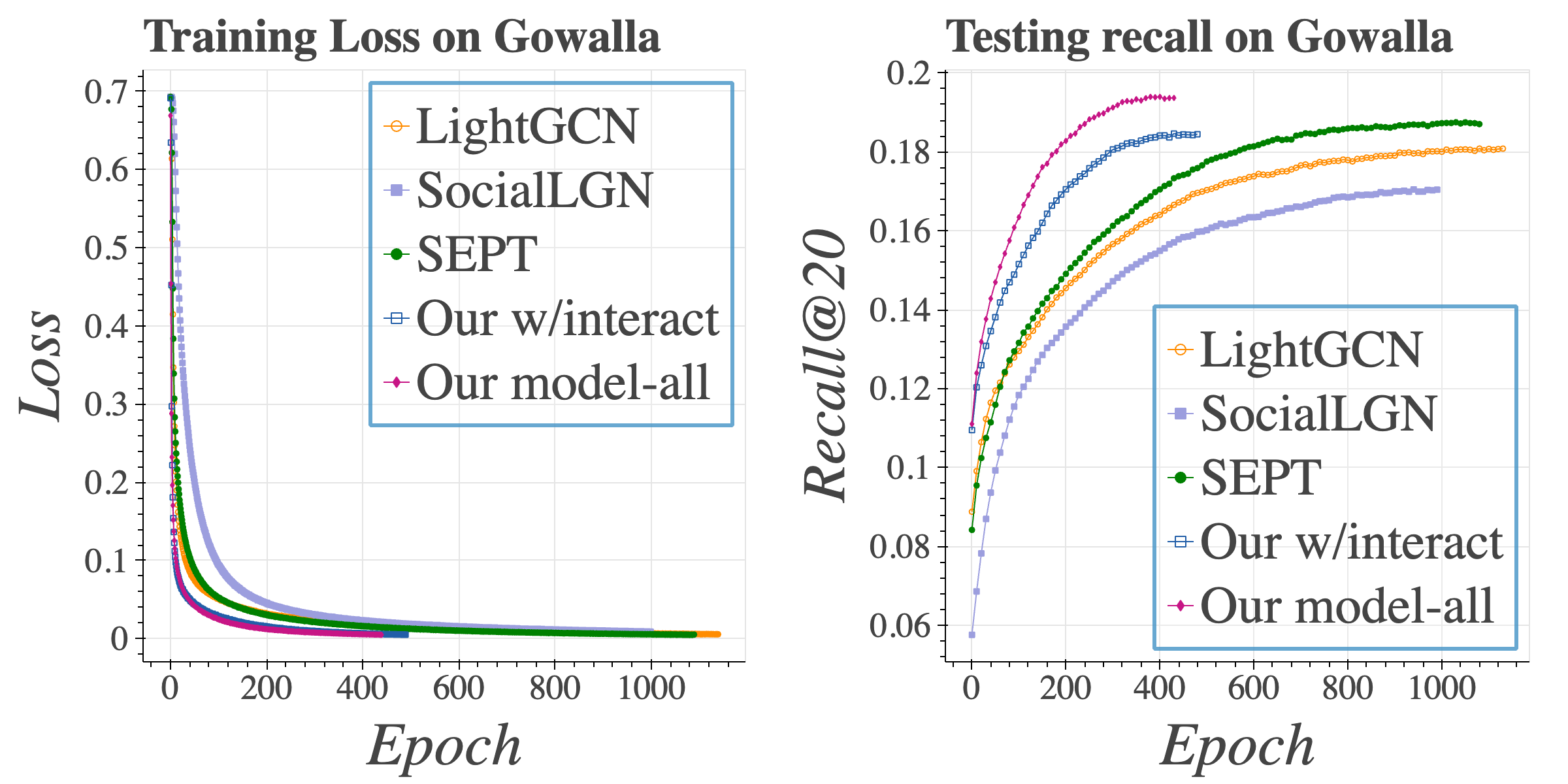}
    \caption{\centering Training curves of LGC-based models, which are evaluated by training loss and testing recall@20 on Gowalla data set.
    \label{fig:gowalla_loss_recall}}
\end{figure}

\begin{figure}[H]
    \centering
    \includegraphics[width = 0.9\linewidth]{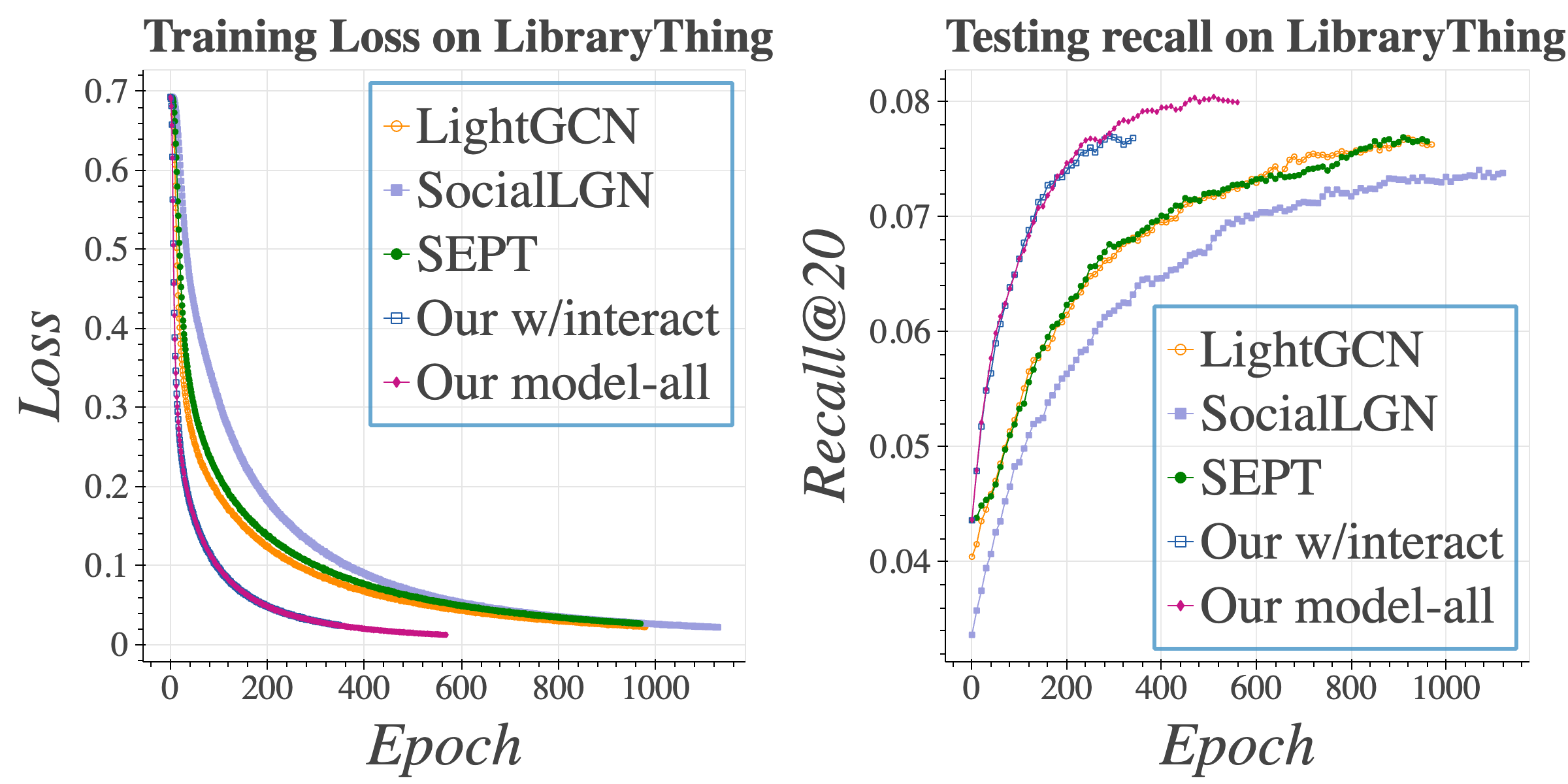}
    \caption{\centering Training curves of LGC-based models, which are evaluated by training loss and testing recall@20 on LibraryThing data set.
\label{fig:librarything_loss_recall}}
\end{figure}

Our model converges faster than the baseline models, shown in the Figure \ref{fig:gowalla_loss_recall} and \ref{fig:librarything_loss_recall}. In our previous work at WiGCN model, adding a weight matrix as input and generating strong attention signals for nodes in the graph during propagation. The user correlation weight matrix based on the set of shared items only needs to be calculated once using basic matrix operations and used for all embedding layers. Therefore, in this article, the user’s correlation matrix with carefully preprocessed $U \times I$ data has created a very high convergence speed for the propagation process.

\subsection{Explain the influence of interaction data and social network friendship data}
One of our main research questions is the influence of interaction data and social friendship data when considered separately and when aggregated. We implemented the model in a modular form, which can enable or disable interactive embedding and social friend embedding, to make our model explainable. In Table \ref{table:influenceimpact} and \ref{table:influenceimpactlibrary}, we compare 4 models: LightGCN (can be considered a baseline model without interaction matrix and social network friendship data); our model only with interaction data or social network friendship data; and our full model.

\begin{table}[H]
\centering
\caption{Influence of interaction data and social network friendship data on Gowalla data set.\label{table:influenceimpact}}
\renewcommand\tabularxcolumn[1]{m{#1}}
\newcolumntype{P}[1]{>{\centering\arraybackslash}p{#1}}
\begin{tabularx}{1.0\linewidth}{|P{2.1cm}|C|C|C|}
\hline
 \textbf{Models} & \textbf{recall} & \textbf{precision} & \textbf{ndcg@20}\\
\hline
LightGCN & 0.1808 &  0.0496 &  0.1488\\
Our w/ interaction &  0.1846 (+2.1\%) &  0.0506 (+2.0\%) &  0.1509 (+1.4\%)\\
Our w/ social &  0.1924 (+6.4\%) &  0.0530 (+6.9\%)&  0.1564 (+5.1\%)\\
Our full model &  0.1940 (+7.3\%) & 0.0537 (+8.3\%)&  0.1572 (+5.6\%)\\
\hline
\end{tabularx}
\end{table}

We conclude that for data sets related to geographical locations on social networking platforms, social friendship data create the main impact. It contributes largely to the increase in precision and recall in our model. For platforms where friendship network is just a secondary function, such as the Librarything data set in Table \ref{table:influenceimpactlibrary}, which have social network just for private messaging, the interaction data plays a crucial role. Finally, when combining both sources of data, the best results are achieved.

\begin{table}[H]
\centering
\caption{Influence of interaction data and social network friendship data on Librarything data set.\label{table:influenceimpactlibrary}}
\renewcommand\tabularxcolumn[1]{m{#1}}
\newcolumntype{P}[1]{>{\centering\arraybackslash}p{#1}}
\begin{tabularx}{1.0\linewidth}{|P{2.1cm}|C|C|C|}
\hline
 \textbf{Models} & \textbf{recall} & \textbf{precision} & \textbf{ndcg@20}\\
\hline
LightGCN & 0.0768&  0.0185&  0.0507\\
Our w/ interaction &  0.0770 (+0.3\%)&  0.0188 (+1.6\%) &  0.0509 (+0.4\%)\\
Our w/ social &  0.0774 (+0.8\%) &  0.0185 (+0\%) &  0.0498 (-1.8\%)\\
Our full model & 0.0804 (+4.7\%)&  0.0191 (+3.2\%) &  0.0533 (+5.1\%)\\
\hline
\end{tabularx}
\end{table}

\section{Conclusion and future works}
\subsection{Conclusion}
In this work, we have presented an effective model that includes data filtering to remove noisy interactions between users and items, evaluating the influence of the weight matrix and activation function on the result, and a method of consolidating data when the similarity of users comes from multiple information sources, as well as comparing the effective recommender system models presented recently \cite{tinvidu}

Designing the input with multi-embedding makes them quickly capture interaction signals and significantly increase the efficiency of the model. Our model has inherited the most effective features from the previously introduced algorithms, which are using the high-order connectivity learning structure, removing the weight matrix, and eliminating the nonlinear activation function. We also evaluated the influence of social friendships on the recommendation process and pointed out some of the difficulties of using this data.

\subsection{Future work}
During the process of conducting research, we proposed several issues that need to be further explored. That is group recommendation, which gives the results not only for a single user but for a whole group of friends. That recommendation model is useful in contexts where a group of friends go to the movies together, go to the same attraction, or a group of students take a course. Another problem is that items have been treated as individuals; they are not clustered or grouped into catalogs. This may have omitted a lot of information from the set of items in the final recommendation result. In fact, items are closely related when they belong to the same commodity group, when items are in geographically distant locations, or when the item is in movies in the same genre.

\begin{IEEEbiography}[{\includegraphics[width=1in,height=1.25in,clip,keepaspectratio]{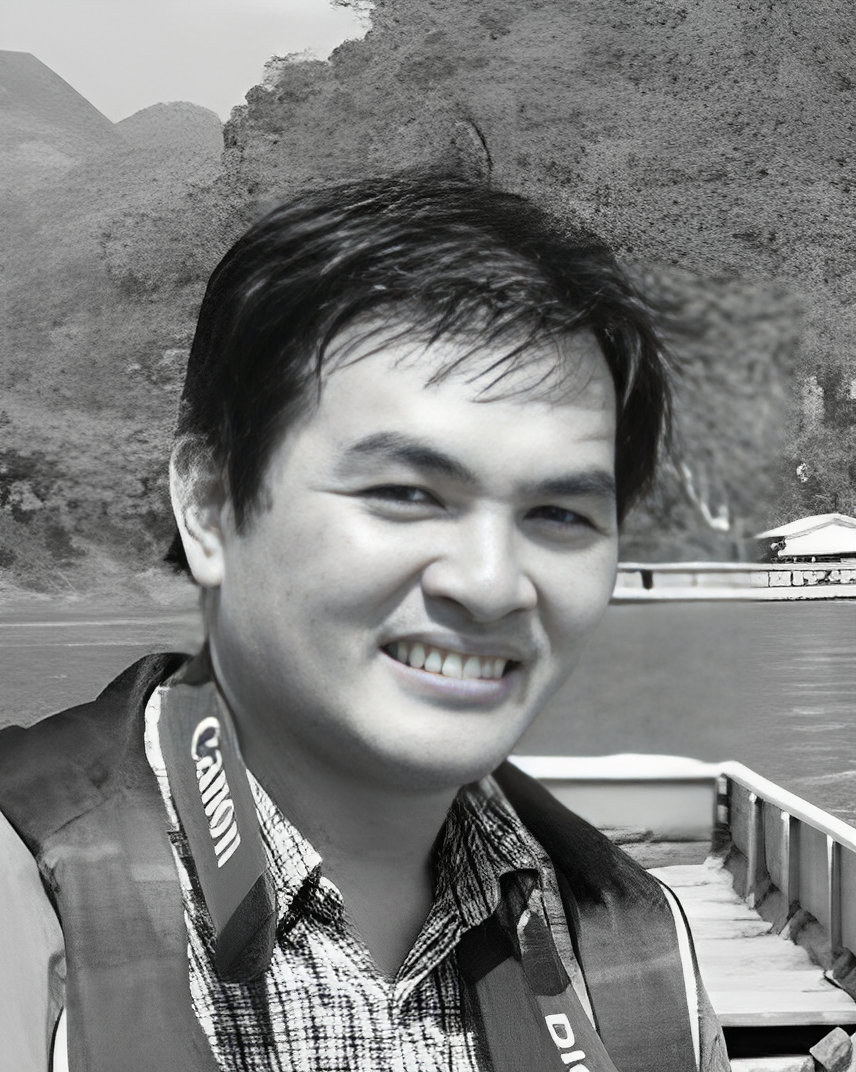}}]{TIN T. TRAN} received the M.S. degree in Computer Science from HCMUT—University of Technology, Ho Chi Minh city, Vietnam in 2012 and is studying a PhD program at VSB—Technical University of Ostrava, Czech Republic. He is working as a Lecturer at the Faculty of Information Technology, Ton Duc Thang university, Vietnam. His research interests include artificial intelligence, social network, graph neural network and location-based application.
\end{IEEEbiography}

\begin{IEEEbiography}[{\includegraphics[width=1in,height=1.25in,clip,keepaspectratio]{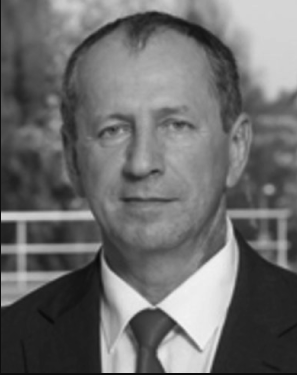}}]{VACLAC SNASEL.} (Senior Member, IEEE) research and development experience includes over 25 years in the Industry and Academia. He works in a multi-disciplinary environment involving artificial intelligence, multidimensional data indexing, conceptual lattice, information retrieval, semantic web, knowledge management, data compression, machine intelligence, neural network, web intelligence, data mining and applied to various real-world problems. He has given more than 10 plenaries lectures and conference tutorials in these areas. He has authored/co-authored several refereed journal/conference papers and book chapters. He has published more than 400 papers (147 are recorded at Web of Science). He has supervised many Ph.D. students from the Czech Republic, Jordan, Yemen,
Slovakia, Ukraine, and Vietnam.  
\end{IEEEbiography}

\begin{IEEEbiography}[{\includegraphics[width=1in,height=1.25in,clip,keepaspectratio]{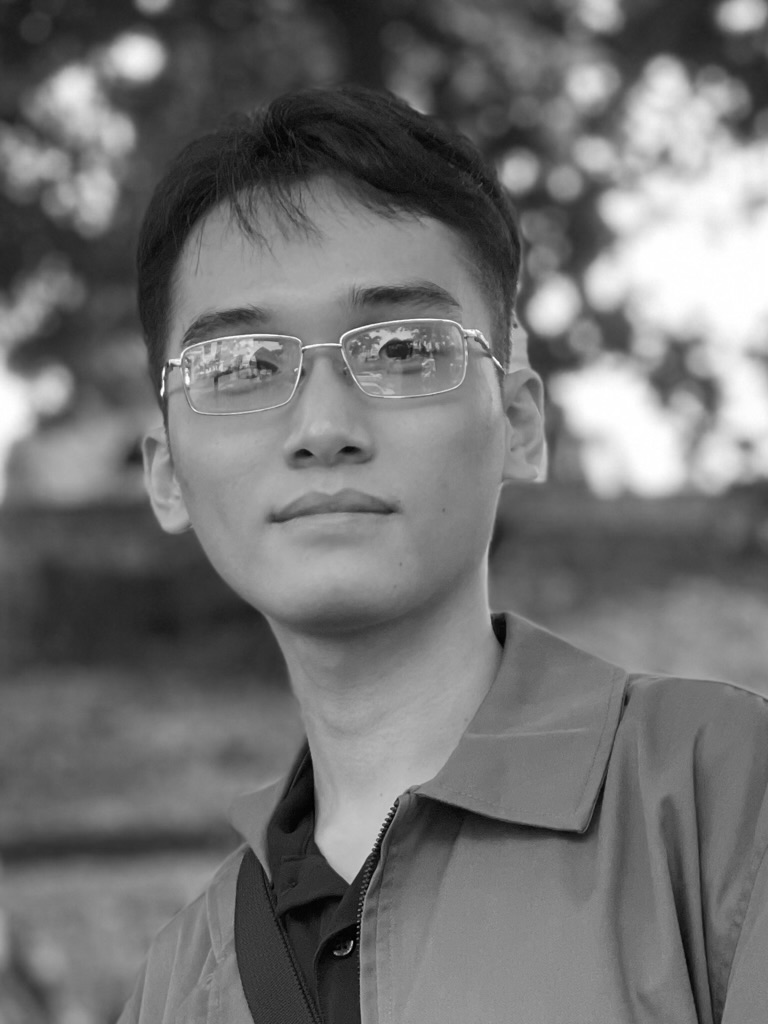}}]{LOC TAN NGUYEN} received the B.S. degree in Computer Science from Ton Duc Thang university, Ho Chi Minh city, Vietnam, in 2023. His research interest includes the data science, data processing, social network, and graph neural network.
\end{IEEEbiography}

\EOD


\begin{thebibliography}{00}
\bibliographystyle{plainnat}
%
\bibitem{5360986}Zhang, J., Lin, Z., Xiao, B. \& Zhang, C. An optimized item-based collaborative filtering recommendation algorithm. {\em 2009 IEEE International Conference On Network Infrastructure And Digital Content}. pp. 414-418 (2009)

\bibitem{6779375}Tewari, A., Kumar, A. \& Barman, A. Book recommendation system based on combine features of content based filtering, collaborative filtering and association rule mining. {\em 2014 IEEE International Advance Computing Conference (IACC)}. pp. 500-503 (2014)
\bibitem{9321429}Grattarola, D. \& Alippi, C. Graph Neural Networks in TensorFlow and Keras with Spektral [Application Notes]. {\em IEEE Computational Intelligence Magazine}. \textbf{16}, 99-106 (2021)
\bibitem{1458205}Ma, H., Yang, H., Lyu, M. \& King, I. SoRec: Social Recommendation Using Probabilistic Matrix Factorization. {\em Proceedings Of The 17th ACM Conference On Information And Knowledge Management}. pp. 931-940 (2008), https://doi.org/10.1145/1458082.1458205
\bibitem{SoReg}Ma, H., Zhou, D., Liu, C., Lyu, M. \& King, I. Recommender systems with social regularization. {\em WSDM}. pp. 287-296 (2011,1)
\bibitem{10068865}Dossena, M., Irwin, C. \& Portinale, L. Graph-based Recommendation using Graph Neural Networks. {\em 2022 21st IEEE International Conference On Machine Learning And Applications (ICMLA)}. pp. 1769-1774 (2022)
\bibitem{NGCF}Wang, X., He, X., Wang, M., Feng, F. \& Chua, T. Neural Graph Collaborative Filtering. {\em CoRR}. \textbf{abs/1905.08108} (2019), http://arxiv.org/abs/1905.08108
\bibitem{Candillier2007ComparingSC}Candillier, L., Meyer, F. \& Boullé, M. Comparing State-of-the-Art Collaborative Filtering Systems. {\em MLDM}. (2007)


\bibitem{Jalili18}Jalili, M., Ahmadian, S., Izadi, M., Moradi, P. \& Salehi, M. Evaluating Collaborative Filtering Recommender Algorithms: A Survey. {\em IEEE Access}. \textbf{6} pp. 74003-74024 (2018)
\bibitem{Pazzani99aframework}Pazzani, M. A Framework for Collaborative, Content-Based and Demographic Filtering. {\em ARTIFICIAL INTELLIGENCE REVIEW}. \textbf{13} pp. 393-408 (1999)
\bibitem{adomavicius2005incorporating}Adomavicius, G., Sankaranarayanan, R., Sen, S. \& Tuzhilin, A. Incorporating contextual information in recommender systems using a multidimensional approach. {\em ACM Transactions On Information Systems (TOIS)}. \textbf{23}, 103-145 (2005)
\bibitem{6313742}Bojnordi, E. \& Moradi, P. A novel collaborative filtering model based on combination of correlation method with matrix completion technique. {\em The 16th CSI International Symposium On Artificial Intelligence And Signal Processing (AISP 2012)}. pp. 191-194 (2012)

\bibitem{10.1145/138859.138867}Goldberg, D., Nichols, D., Oki, B. \& Terry, D. Using Collaborative Filtering to Weave an Information Tapestry. {\em Commun. ACM}. \textbf{35}, 61-70 (1992,12), https://doi.org/10.1145/138859.138867
\bibitem{doi:10.1137/1.9781611972726.4}Yu, K., Xu, X., Tao, J., Ester, M. \& Kriegel, H. Instance Selection Techniques for Memory-Based Collaborative Filtering. {\em Proceedings Of The 2002 SIAM International Conference On Data Mining (SDM)}. pp. 59-74, https://epubs.siam.org/doi/abs/10.1137/1.9781611972726.4
\bibitem{5197422}Koren, Y., Bell, R. \& Volinsky, C. Matrix Factorization Techniques for Recommender Systems. {\em Computer}. \textbf{42}, 30-37 (2009)
\bibitem{berg2017graph}Berg, R., Kipf, T. \& Welling, M. Graph Convolutional Matrix Completion.  (2017)

\bibitem{9216015}Guo, Q., Zhuang, F., Qin, C., Zhu, H., Xie, X., Xiong, H. \& He, Q. A Survey on Knowledge Graph-Based Recommender Systems. {\em IEEE Transactions On Knowledge; Data Engineering}. \textbf{34}, 3549-3568 (2022,8)
\bibitem{gao2020deep}Gao, Y., Li, Y., Lin, Y., Gao, H. \& Khan, L. Deep Learning on Knowledge Graph for Recommender System: A Survey.  (2020)

\bibitem{wu2021graph}Wu, S., Sun, F., Zhang, W. \& Cui, B. Graph Neural Networks in Recommender Systems: A Survey.  (2021)
\bibitem{Wu2021}Wu, Z., Pan, S., Chen, F., Long, G., Zhang, C. \& Yu, P. A Comprehensive Survey on Graph Neural Networks. {\em IEEE Transactions On Neural Networks And Learning Systems}. \textbf{32}, 4-24 (2021,1), http://dx.doi.org/10.1109/TNNLS.2020.2978386
\bibitem{DBLP:journals/corr/abs-1903-07293}Wang, X., Ji, H., Shi, C., Wang, B., Cui, P., Yu, P. \& Ye, Y. Heterogeneous Graph Attention Network. {\em CoRR}. \textbf{abs/1903.07293} (2019), http://arxiv.org/abs/1903.07293
\bibitem{10.1145/3219819.3219890}Ying, R., He, R., Chen, K., Eksombatchai, P., Hamilton, W. \& Leskovec, J. Graph Convolutional Neural Networks for Web-Scale Recommender Systems. {\em Proceedings Of The 24th ACM SIGKDD International Conference On Knowledge Discovery \&; Data Mining}. pp. 974-983 (2018), https://doi.org/10.1145/3219819.3219890
\bibitem{DBLP:journals/corr/HamiltonYL17}Hamilton, W., Ying, R. \& Leskovec, J. Inductive Representation Learning on Large Graphs. {\em CoRR}. \textbf{abs/1706.02216} (2017), http://arxiv.org/abs/1706.02216
\bibitem{he2020lightgcn}He, X., Deng, K., Wang, X., Li, Y., Zhang, Y. \& Wang, M. LightGCN: Simplifying and Powering Graph Convolution Network for Recommendation.  (2020)

%
%
\bibitem{7404241}Guo, G., Zhang, J. \& Yorke-Smith, N. A Novel Recommendation Model Regularized with User Trust and Item Ratings. {\em IEEE Transactions On Knowledge And Data Engineering}. \textbf{28}, 1607-1620 (2016)
\bibitem{6714549}Jiang, M., Cui, P., Wang, F., Zhu, W. \& Yang, S. Scalable Recommendation with Social Contextual Information. {\em IEEE Transactions On Knowledge And Data Engineering}. \textbf{26}, 2789-2802 (2014)
\bibitem{Bond2012}Bond, R., Fariss, C., Jones, J., Kramer, A., Marlow, C., Settle, J. \& Fowler, J. A 61-million-person experiment in social influence and political mobilization. {\em Nature}. \textbf{489}, 295-298 (2012,9,1), https://doi.org/10.1038/nature11421
\bibitem{Qiu2018DeepInfM}Qiu, J., Tang, J., Ma, H., Dong, Y., Wang, K. \& Tang, J. DeepInf : Modeling Influence Locality in Large Social Networks.  (2018), https://api.semanticscholar.org/CorpusID:49239721
\bibitem{Wang2018ContextMFA}Wang, J., Bagul, D. \& Srihari, S. ContextMF : A Fast and Context-aware Embedding Learning Method for Recommendation Systems.  (2018), https://api.semanticscholar.org/CorpusID:198186939
\bibitem{10.1145/1401890.1401944}Koren, Y. Factorization Meets the Neighborhood: A Multifaceted Collaborative Filtering Model. {\em Proceedings Of The 14th ACM SIGKDD International Conference On Knowledge Discovery And Data Mining}. pp. 426-434 (2008), https://doi.org/10.1145/1401890.1401944
\bibitem{fan2019graph}Fan, W., Ma, Y., Li, Q., He, Y., Zhao, E., Tang, J. \& Yin, D. Graph Neural Networks for Social Recommendation.  (2019)
\bibitem{LIAO2022595}Liao, J., Zhou, W., Luo, F., Wen, J., Gao, M., Li, X. \& Zeng, J. SocialLGN: Light graph convolution network for social recommendation. {\em Information Sciences}. \textbf{589} pp. 595-607 (2022), https://www.sciencedirect.com/science/article/pii/S0020025522000019 

\bibitem{DBLP:journals/corr/abs-2106-03569}Yu, J., Yin, H., Gao, M., Xia, X., Zhang, X. \& Hung, N. Socially-Aware Self-Supervised Tri-Training for Recommendation. {\em CoRR}. \textbf{abs/2106.03569} (2021), https://arxiv.org/abs/2106.03569



%

\bibitem{wigcn}Tran, T. \& Snasel, V. Improvement Graph Convolution Collaborative Filtering with Weighted Addition Input. {\em Intelligent Information And Database Systems}. pp. 635-647 (2022)

\bibitem{najork2008computing}Najork, M. \& McSherry, F. Computing Information Retrieval Performance Measures Efficiently in the Presence of Tied Scores. {\em 30th European Conference On IR Research (ECIR)}. (2008,4), https://www.microsoft.com/en-us/research/publication/computing-information-retrieval-performance-measures-efficiently-in-the-presence-of-tied-scores/

\bibitem{he2017neural}He, X., Liao, L., Zhang, H., Nie, L., Hu, X. \& Chua, T. Neural Collaborative Filtering.  (2017)





\bibitem{rendle2012bpr}Rendle, S., Freudenthaler, C., Gantner, Z. \& Schmidt-Thieme, L. BPR: Bayesian Personalized Ranking from Implicit Feedback.  (2012)

\bibitem{kingma2017adam}Kingma, D. \& Ba, J. Adam: A Method for Stochastic Optimization.  (2017)




%
%
%
%






\bibitem{10.1145/2020408.2020579}Cho, E., Myers, S. \& Leskovec, J. Friendship and Mobility: User Movement in Location-Based Social Networks. {\em Proceedings Of The 17th ACM SIGKDD International Conference On Knowledge Discovery And Data Mining}. pp. 1082-1090 (2011), https://doi.org/10.1145/2020408.2020579

\bibitem{10.1145/2806416.2806511}Zhao, T., McAuley, J. \& King, I. Improving Latent Factor Models via Personalized Feature Projection for One Class Recommendation. {\em Proceedings Of The 24th ACM International On Conference On Information And Knowledge Management}. pp. 821-830 (2015), https://doi.org/10.1145/2806416.2806511
\bibitem{10.1145/2124295.2124309}Tang, J., Gao, H. \& Liu, H. MTrust: Discerning Multi-Faceted Trust in a Connected World. {\em Proceedings Of The Fifth ACM International Conference On Web Search And Data Mining}. pp. 93-102 (2012), https://doi.org/10.1145/2124295.2124309
\bibitem{epinionds}Reyhani Hamedani, M., Ali, I., Hong, J. \& Kim, S. TrustRec: An effective approach to exploit implicit trust and distrust relationships along with explicit ones for accurate recommendations. {\em Computer Science And Information Systems}. \textbf{18} pp. 93-114 (2021,1)
\bibitem{10.1145/963770.963772}Herlocker, J., Konstan, J., Terveen, L. \& Riedl, J. Evaluating Collaborative Filtering Recommender Systems. {\em ACM Trans. Inf. Syst.}. \textbf{22}, 5-53 (2004,1), https://doi.org/10.1145/963770.963772

\bibitem{Sarwar00}Sarwar, B., Karypis, G., Konstan, J. \& Riedl, J. Application of Dimensionality Reduction in Recommender System – A Case Study.  (2000,8)

\bibitem{Sarwar00E}Sarwar, B., Karypis, G., Konstan, J. \& Riedl, J. Analysis of Recommendation Algorithms for E-Commerce. {\em Proceedings Of The 2nd ACM Conference On Electronic Commerce}. pp. 158-167 (2000), https://doi.org/10.1145/352871.352887

\bibitem{10.1145/1864708.1864736}Jamali, M. \& Ester, M. A Matrix Factorization Technique with Trust Propagation for Recommendation in Social Networks. {\em Proceedings Of The Fourth ACM Conference On Recommender Systems}. pp. 135-142 (2010), https://doi.org/10.1145/1864708.1864736

\end{thebibliography}
\end{document}